\documentclass[aps,prl,reprint,superscriptaddress,letterpaper,showpacs,twocolumn,citeautoscript]{revtex4-1}
\usepackage[english]{babel}
\usepackage{graphicx}
\usepackage[pdftex                     
           ,pagebackref=false          
           ,colorlinks=false           
           ]{hyperref}
\hypersetup{linkcolor=blue,            
            citecolor=red,             
            urlcolor=black}            

\usepackage{amsmath}
\usepackage{mathtools}
\usepackage{eufrak}
\usepackage{times}

\newcommand*{\abinitio}{{\itshape ab initio}}
\newcommand*{\fleur}{\texttt{FLEUR}}

\newcommand*{\DM}{Dzyaloshinskii-Moriya}
\newcommand*{\vn}{\boldsymbol}

\begin{document}

\begin{abstract}
We address the importance of the modern theory of orbital magnetization for spintronics. Based on an all-electron first-principles approach, we demonstrate that the predictive power of the routinely
employed ``atom-centered" approximation is limited to materials like elemental bulk ferromagnets, while the application of the 
modern theory of orbital magnetization is crucial in chemically or structurally inhomogeneous systems such as 
magnetic thin films, and materials exhibiting nontrivial topology in reciprocal and real space,~e.g.,~Chern 
insulators or noncollinear systems. We find that the modern theory is particularly 
crucial for describing magnetism in a class of materials that we suggest here---topological orbital ferromagnets.
\end{abstract}

\setcounter{secnumdepth}{3}
 \title{
Role of Berry phase theory for describing orbital magnetism:
From magnetic heterostructures to topological orbital ferromagnets}
 \author{J.-P. Hanke}
 \email{j.hanke@fz-juelich.de}
 \author{F. Freimuth}
 \author{A. K. Nandy}
  \affiliation{Peter Gr\"unberg Institut and Institute for Advanced Simulation,\\Forschungszentrum J\"ulich and JARA, 52425 J\"ulich, Germany}
 \author{H. Zhang}
 \affiliation{TU Darmstadt, Alarich-Weiss-Stra{\ss}e 2, 64287 Darmstadt, Germany}
 \author{S. Bl\"ugel}
 \author{Y. Mokrousov}
 \affiliation{Peter Gr\"unberg Institut and Institute for Advanced Simulation,\\Forschungszentrum J\"ulich and JARA, 52425 J\"ulich, Germany}
 \maketitle

Magnetism is an elementary property of materials, and it is composed of spin and orbital contributions. In contrast to the concept of spin magnetization, which has been relatively well understood and extensively researched in the course of the past decades, our understanding of orbital magnetism in solids has been poor so far, and an ability to describe it reliably has been missing until recently. Both spin and orbital magnetization (OM) are accessible separately, e.g., by means of magnetomechanical~\cite{Meyer1961} or magnetic circular dichroism measurements~\cite{Thole1992,Carra1993,Chen1995}, but the orbital contribution to the magnetization of solids is usually overshadowed by the spin counterpart, owing to the orbital moment quenching. However, in certain systems the OM yields an equally important contribution, which can even result in a spin-orbital compensation of magnetization~\cite{Taylor2002,Gotsis2003,Qiao2004}. Its influence on spin-dependent transport~\cite{Xiao2006,Murakami2006,Xiao2007,Wang2007}, magnetic susceptibility~\cite{Wang2007}, orbital magnetoelectric response~\cite{Malashevich2010,Essin2009,Essin2010}, magnetic anisotropy~\cite{Vleck1937}, and \DM{} interaction~\cite{Moriya1960} renders the OM crucial for understanding basic properties of magnets.
A spontaneous OM in ferromagnets is a key manifestation of the spin-orbit interaction (SOI), lifting in part the quenching mechanism. This interpretation applies to most materials but it fails to explain orbital magnetism in systems where a finite {\it topological} OM emerges even without SOI as a result of a nontrivial real-space distribution of spins~\cite{Hoffmann2015}.

Addressing the OM in solids is a subtle point as the position operator $\vn r$ is ill-defined in the basis of extended Bloch states. To circumvent this problem in \abinitio{} calculations, the evaluation of the angular momentum operator $\vn L$ is typically restricted locally in space to atom-centered spheres. This atom-centered approximation (ACA) is widely used to study orbital magnetism in solids even nowadays. Rather recently, a rigorous theory of OM was established through three independent approaches~\cite{Thonhauser2005,Ceresoli2006,Xiao2005,Shi2007}. In this so-called {\it modern theory}~\cite{Thonhauser2011,Resta2010,Xiao2010} the OM is expressed as a genuine bulk property evaluated from the ground-state wave functions:
\begin{equation}
 \vn m = \frac{e}{2\hbar}\, \text{Im} \int[dk]\, \langle \partial_{\vn k} u_{\vn kn}| \times\left(H_{\vn k} + \mathcal E_{\vn kn} - 2 \mathcal E_F\right)| \partial_{\vn k} u_{\vn kn}\rangle \, ,
 \label{eq:def_orbmom}
\end{equation}
where $\vn k$ is the crystal momentum, $[d k]$ stands for $\sum_n^{\text{occ}}\text d\vn k /(2\pi)^3$, $|u_{\vn kn}\rangle$ is an eigenstate of the lattice-periodic Hamiltonian $H_{\vn k}= e^{-i\vn k \cdot \vn r} H e^{i \vn k \cdot \vn r}$ to the band energy $\mathcal E_{\vn kn}$, $\mathcal E_F$ is the Fermi energy, and $e>0$ is the elementary positive charge. At zero temperature, the summation is restricted to all occupied bands $n$ below the Fermi energy. In contrast to the ACA, Eq.~\eqref{eq:def_orbmom} naturally and unambiguously accounts for nonlocal contributions to the OM~\cite{Nikolaev2014}.

How important is the modern framework for accessing the orbital magnetism in 
systems which are of great interest in today's spintronics,\,e.g., metallic magnetic thin films~\cite{Miron2011}, topologically nontrivial materials
such as Chern insulators~\cite{Chang2013}, or magnetically complex systems such as frustrated spin lattices and skyrmions~\cite{Taguchi2001,Heinze2011}?  Although the modern theory of OM has already been implemented in several first-principles electronic-structure codes based on pseudopotentials, its comparison to the widely used and computationally cheap ACA applied in all-electron methods has been performed only for bulk Fe, Co, Ni, and several perovskite transition-metal oxides~\cite{Ceresoli2010,Lopez2012,Nikolaev2014}. And while in the latter case the modern theory does not significantly alter the values of the OM, the ACA was found to underestimate the OM in bcc Fe by up to $50$\,\%. However, the
relatively modest magnitude of the OM in elemental Fe leaves open the question of relevance of the modern theory in 
wide classes of materials explored today. In particular, until now, a strong justification for the computationally very challenging modern-theory description by all-electron approaches is missing, leading to significant doubts on the wider relevance of the modern theory of OM in magnetic systems.

Here, based on first-principles calculations we evaluate the importance of the modern theory of OM in elemental bulk ferromagnets, magnetic thin films, Chern insulators, and systems with non-collinear magnetism. We do this by contrasting the ACA with the modern theory of OM in these systems, Eq.~\eqref{eq:def_orbmom}. We demonstrate that using the modern theory is essential for complex magnetic materials, and while ACA performs well in
elemental bulk ferromagnets, it breaks down completely in a chemically and structurally heterogeneous system of a Mn monolayer deposited on W(001). Furthermore, the correct description of the OM in the Chern insulating phase, modeled here by a system of graphene decorated by $5d$ transition-metal adatoms, requires the modern theory. Finally, we demonstrate the crucial role that the modern theory plays for correctly describing the magnetism in 
the class of materials the discovery of which we display in this work---{\it topological orbital ferromagnets}. In these materials, 
exemplified here by the $3Q$ state of Mn/Cu(111), the macroscopic spin magnetization is completely 
replaced by its orbital counterpart, prominent even without spin-orbit interaction. 

Within  a common ansatz, the orbital moment in the unit cell (uc) $\vn m^{\rm{uc}}_{\vn k n} = -\frac{e}{2m_e} \langle \psi_{\vn k n} | \vn L | \psi_{\vn k n} \rangle$ associated with a state $|\psi_{\vn k n}\rangle$ is obtained by integrating the angular momentum operator over the unit cell. Here, $m_e$ is the electron mass. In all-electron methods the space is often partitioned into muffin-tin spheres centered around the atoms and the interstitial region between the atoms. Typically, the evaluation of the orbital moment is given by the ACA, $\vn{m}^{\mu}_{\vn k n} = -\frac{e}{2m_e} \langle \psi_{\vn k n} | \vn L^{\mu} | \psi_{\vn k n} \rangle_{\text{MT}}$, i.e., only by the contribution of the muffin-tin (MT) sphere of atom $\mu$. Here, $\vn L^{\mu}=\vn r_\mu \times \vn p$ where $\vn r_\mu$ is the position operator with respect to the center of atom $\mu$ and $\vn p$ is the momentum operator. The OM in ACA is computed by summing up the individual contributions over all occupied states and atoms in the unit cell, and dividing by the unit-cell volume $\Omega$:
\begin{equation}
\vn{m} = \frac{1}{N_{\vn k}\Omega}\sum_{\vn k}\sum_n^{\text{occ}} \sum_{\mu}\vn{m}^{\mu}_{\vn k n} \, ,
\label{eq:aca_orbmom}
\end{equation}
where $N_{\vn k}$ is the number of $\vn k$ points. In this work, Eq.~\eqref{eq:aca_orbmom} is contrasted with its modern-theory counterpart, Eq.~\eqref{eq:def_orbmom}. 
To converge efficiently the $\vn k$ summation in the expressions for $\vn{m}$ in Eqs.~\eqref{eq:def_orbmom} and~\eqref{eq:aca_orbmom}, we employ the Wannier interpolation technique~\cite{Yates2007,Lopez2012,suppl}, realized within the accurate all-electron full-potential linearized augmented plane-wave (FLAPW)  
code \fleur{}~\cite{fleur}. Using this code, we perform self-consistent density functional theory calculations including SOI in second variation and using the PBE functional, unless stated otherwise~\cite{Marzari1997,Souza2001,Marzari2012,Mostofi2008,Freimuth2008,suppl}.

\begin{figure}
\centering
\includegraphics{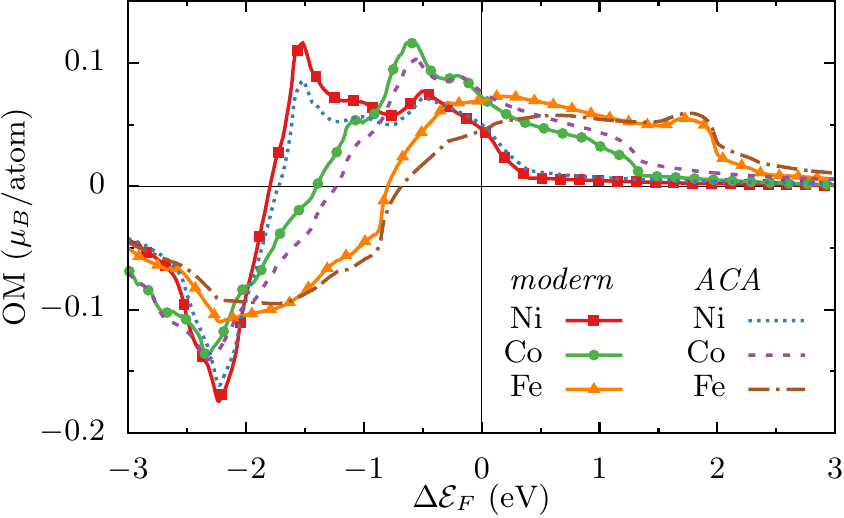}
\caption{Easy-axis orbital magnetization (OM) in the bulk ferromagnets fcc Ni, hcp Co, and bcc Fe, according to atom-centered approximation (ACA) and modern theory (per atom). The Fermi level is varied by $\Delta\mathcal E_F$ with respect to the true Fermi energy.} 
\label{fig:bulk_overview}
\end{figure}

{\it Bulk ferromagnets.} We begin by considering the bulk ferromagnets bcc Fe, hcp Co, and fcc Ni. The magnetization direction is aligned along the experimental easy axis, which is $(001)$ for Fe, $(0001)$ for Co, and $(111)$ for Ni. In Fig.~\ref{fig:bulk_overview} we present the OM as a function of the Fermi energy $\mathcal E_F$ with respect to the true Fermi energy $\mathcal E_F^0$, $\Delta \mathcal E_F = \mathcal E_F - \mathcal E_F^0$. At $\Delta\mathcal E_F=0$, we obtain in ACA the values $0.0451\mu_B/\text{Fe}$, $0.0767\mu_B/\text{Co}$, $0.0499\mu_B/\text{Ni}$, and $0.0693$, $0.0727$, $0.0460$ in the modern theory. The latter results agree well with previous pseudopotential calculations ($0.0761$, $0.0838$, $0.0467$)~\cite{Lopez2012}. Clearly, the agreement between the modern theory and ACA is good for Ni and Co over the whole range of energies. However, in the case of Fe the modern theory corrects the OM by more than $50$\% around the Fermi level, which could be attributed to a larger
degree of delocalization of the Bloch states in this material as compared to Co and Ni. Compared to experiment ($0.081$, $0.133$, $0.053$)~\cite{Meyer1961}, the modern theory particularly improves the OM value in Fe.

{\it Heterogeneous systems.} In heterogeneous materials like thin magnetic films, the local  moments contributing to the OM in ACA can vary strongly in real space, and compensate each other. Therefore, nonlocal effects are expected to play a significant role for the OM in these systems. To prove this point, as an example, we consider an asymmetric slab of a Mn monolayer deposited on nine atomic layers of bcc W, Mn/W(001). The structural parameters taken from Ref.~\cite{Ferriani2005} were adopted for the magnetic interface. Although Mn/W(001) exhibits in reality a long-wavelength spin-spiral ground state~\cite{Ferriani2008,*Ferriani2009}, the collinear ferromagnetic case is studied here.

\begin{figure}
\centering
\includegraphics{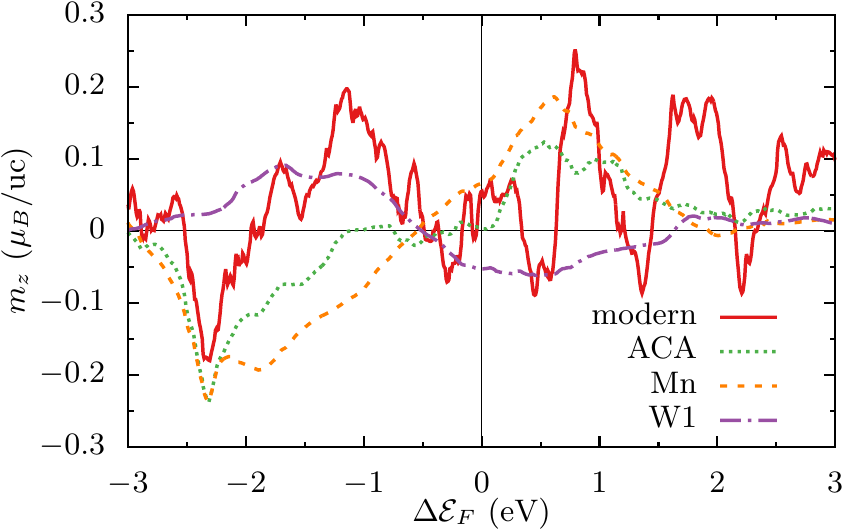}
\caption{Orbital magnetization $m_z$ in Mn/W(001) according to ACA and modern theory (per two-dimensional unit cell, uc). Additionally, the local orbital moments in ACA of Mn and the first W (W1) layer are shown.}
\label{fig:MnW_mom}
\end{figure}

Our first-principles results, presented in Fig.~\ref{fig:MnW_mom}, reveal a drastic difference between the 
modern theory and ACA for the out-of-plane OM ($m_z$). Not only does the modern theory alter the magnitude of the OM, but even its sign is
different from that obtained in ACA over wide regions of energy. The ACA moment is dominated by the local atomic moments in the first two layers, i.e., the Mn overlayer and the first layer of W
atoms (W1). These moments compensate partly leading to an underestimation of orbital magnetism. Noticeably, the effect is pronounced near the Fermi level where the OM is drastically reduced by an order of magnitude as compared to its modern-theory value. Since magnetism originates primarily in Mn, while only W exhibits strong SOI, nonlocal effects become important in this material. As a consequence, the ACA performs particularly poorly with respect to the modern theory. The manifestly nontrivial, rapidly oscillating behavior of the modern theory OM with energy, 
typical for properties driven by the Berry curvature in reciprocal space, manifests the complexity of the orbital
magnetism in magnetic thin films, and calls for revisiting our understanding of orbital physics at surfaces. 

{\it Chern insulators.} Here we test the importance of the modern theory for OM in realistic systems which exhibit topologically nontrivial gaps in their spectrum. Previous work has shown that $5d$ transition-metal adatoms deposited on graphene support strong magnetoelectric response and Chern insulator band gaps due to SOI~\cite{Zhang2012}. As an example, we consider the 
system of ferromagnetically coupled W adatoms with a spin moment of $1.6\mu_B$ deposited on graphene in a $4\times 4$ geometry. W is placed at the hollow sites of free-standing graphene, with the magnetization out-of-plane (along the $z$-axis)~\footnote{The atomic coordinates and computational parameters from Ref.~\cite{Zhang2012} were used}. Upon considering SOI, as a consequence of complex hybridization between the $d$ states of W and graphene $p$ states, a global band gap opens directly at the Fermi level and approximately $0.27$ eV below it. Due to the topologically nontrivial nature of these gaps, the Chern number $\mathcal C_1 = \frac{1}{2\pi}\int \Omega_{xy} d\vn k$ takes the quantized values of $+2$ and $-2$, respectively, and the anomalous Hall conductivity (AHC) is $\sigma_{xy}=-\frac{e^2}{h}\mathcal C_1$. Here, $\Omega_{xy}=-2\text{Im}\sum_n^{\text{occ}}\langle \partial_{k_x}u_{\vn k n}|\partial_{k_y}u_{\vn k n}\rangle$ is the only nonvanishing (in two dimensions) component of the Berry curvature tensor of all occupied states below the respective gap. It follows from Eq.~\eqref{eq:def_orbmom} that $\frac{d m_z}{d \mathcal E_F} =  \frac{e}{h}\mathcal C_1$ in the Chern insulator phase~\cite{Thonhauser2011}.

\begin{figure}
\centering
\includegraphics{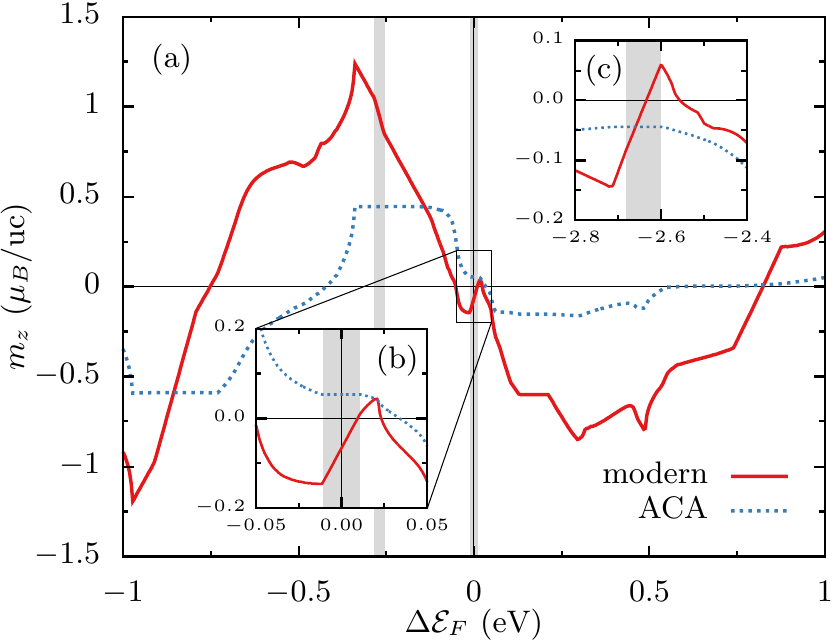}
\caption{(a) Orbital magnetization $m_z$ according to modern theory and atom-centered approximation (ACA) for the W-graphene hybrid system. The shaded regions highlight nontrivial Chern insulator gaps. (b) Zoom to the region near the Fermi level. (c) Orbital magnetization for Ir deposited in $2\times2$ geometry on graphene. A nontrivial band gap with the Chern number $+2$ opens about $2.65$\,eV below the Fermi level.}
\label{fig:IrGr_mom}
\end{figure}

In Fig.~\ref{fig:IrGr_mom} the performance of the modern theory with respect to ACA is presented. By inspecting the shaded regions of the topologically nontrivial gaps in this figure, we observe that the modern theory OM is perfectly linear in $\Delta\mathcal E_F$ as expected, and even changes its sign around the Fermi level [cf. Fig.~\ref{fig:IrGr_mom}(b)]. This is in sharp contrast to the ACA, which predicts a constant value of OM within the gaps. Replacing W with other $5d$ transition metals, for example Ir, we observe the same breakdown of the ACA in the vicinity of the Chern insulator gaps in the spectrum; see, e.g.,~Fig.~\ref{fig:IrGr_mom}(c). Despite
the fact that both pronounced spin magnetism and strong SOI originate from the same atomic species (W),
the overall agreement of modern theory with ACA is very poor not only directly within the Chern insulator gaps, but also in a wider region around them,~Fig.~\ref{fig:IrGr_mom}(a). This can be understood from the observation of strong interaction between graphene and W 
states, which at the end leads to the formation of topologically nontrivial gaps. Finally, we remark that the energy dependence of the OM and the AHC are not overall correlated in this system.

\begin{figure}
\centering
\scalebox{0.98}{\includegraphics{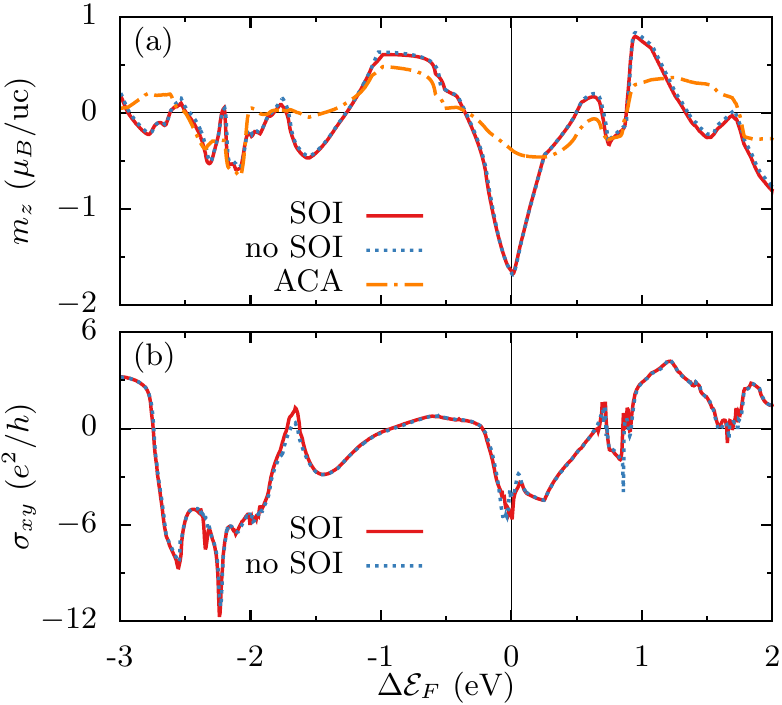}}\vspace{0.4cm}
\scalebox{0.080}{\includegraphics{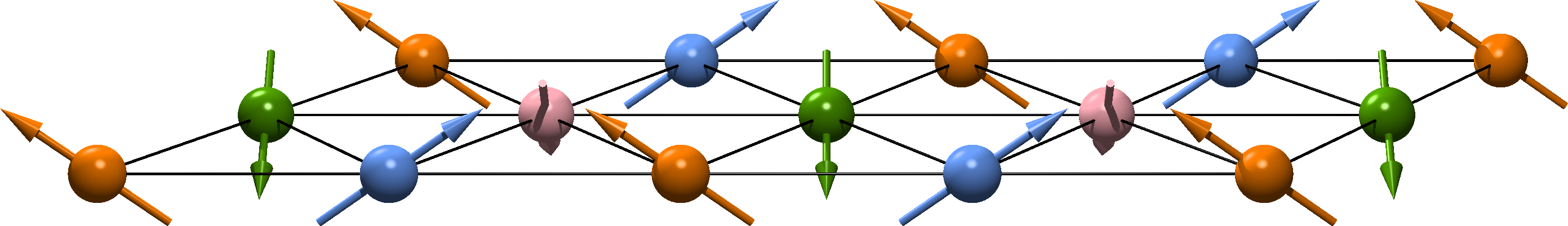}}
\caption{Top: (a) Orbital magnetization $m_z$ and (b) anomalous Hall conductivity $\sigma_{xy}$ of the unsupported Mn monolayer in $3Q$ state, with and without SOI. The dash-dotted line refers to ACA
values of the OM. Bottom: Three-dimensional $3Q$ state of the unsupported Mn monolayer.}
\label{fig:UML_3Q}
\end{figure}

{\it Topological orbital ferromagnets.} The competing exchange interactions between itinerant spins on the two-dimensional triangular lattice can realize noncollinear magnetic structures. A prime example is the superposition of three spiral spin-density waves (SSDWs) with finite wave vectors $\vn Q^{(i)}$, $i=1,2,3$. This so-called $3Q$ state (cf. Fig.~\ref{fig:UML_3Q}) exhibits no net spin magnetization and it is the ground-state spin structure of a Mn monolayer deposited on Cu(111), Mn/Cu(111)~\cite{Kurz2001}, for which we study the orbital magnetism here.

In contrast to typical interfacial systems, which often exhibit chiral magnetic states due to SOI-mediated Dzyaloshinskii-Moriya interaction~\cite{Dzyaloshinsky1958,Moriya1960}, the $3Q$ state of Mn/Cu(111) is a result of competing isotropic higher-order exchange interactions, and it is practically not altered
upon considering SOI. The total spin magnetization in the unit cell is zero. Assuming locally a collinear alignment of orbital and spin moment in the presence of SOI, also the total OM is expected to be zero. We show below, however, that this is not the case. Since the electronic structure of Cu around the Fermi level is dominated by $s$ electrons and the $3d$-$3d$ hybridization between the overlayer and the substrate is small, we modeled the system as an unsupported Mn(111) monolayer at the lattice constant of Cu(111). In Fig.~\ref{fig:UML_3Q} we present our results for the out-of-plane ($z$) component of the OM which is the only nonvanishing one. Strikingly, we observe that the OM does not follow the direction of the spin moment, but is determined by the symmetry of the film, and that the ACA serves as a very crude approximation to the OM, resulting in large differences when compared to the modern-theory values. In particular near the Fermi level, a large underestimation of the OM by the ACA is apparent, with the ACA giving rise to an OM at the Fermi energy which is four times smaller than the modern-theory value.
We also observe that the energy dependence of the OM is correlated much stronger with that of the AHC, Fig.~\ref{fig:UML_3Q}, as opposed to the cases discussed above.

The SOI is well known to be important for OM and AHC. Strikingly, we find that both properties do not rely on the presence of SOI but stem manifestly from the noncollinear spin texture of the $3Q$ state. However, as opposed to the case of the spin lattice of Fe/Ir(001)~\cite{Hoffmann2015} for which a contribution to the OM without SOI has been also observed, in Mn/Cu(111) the presence of SOI makes no noticeable effect on OM and AHC. The AHE in this case can be seen as a purely {\it topological} Hall effect \cite{Hoffmann2015,Franz2014,Kuebler2014}. Remarkably, the large overall magnetization of about $-1.5\mu_B$ per unit cell of Mn/Cu(111) at the true Fermi energy is entirely due to orbital magnetism. The system of Mn/Cu(111) is thus a representative of a class of materials, which we refer to as {\it topological orbital ferromagnets} (TOFs),~i.e., ferromagnets for
which the macroscopic magnetization is solely dominated by the OM, with the latter originating from the nontrivial topology of spin distribution in real space, rather than SOI. 

The origin of the latter {\it topological} OM (TOM) can be attributed to the presence of an ``emergent" magnetic field, which roots in the non-coplanarity of the neighboring spins~\cite{Shindou2001}, and which plays also a crucial role in the physics of skyrmions~\cite{Bruno2004,Neubauer2009,Kanazawa2011,Huang2012,Franz2014,Gayles2015}. The emergent field couples to the orbital degrees of freedom and is identified as an alternative mechanism lifting the orbital degeneracy~\cite{Hoffmann2015}. As TOM is a consequence of the complex noncollinear structure of the delocalized Bloch wave functions in real space, the importance of the nonlocal contributions in this case calls for a proper modern-theory description of orbital magnetism in noncollinear structures such as multi-$Q$ states and skyrmions.

The emergence of a ferromagnetic ordering of large TOM as we predict in these zero-spin-magnetization magnets opens a path to intriguing physics as orbital moments couple to external magnetic fields, optical perturbations, and orbital currents. For example, it is known that the chiral correlation between the spins on a lattice can display high stability with respect to fluctuations (see, e.g.,~\cite{Menzel2012}), and we speculate that the long-range ferromagnetism of TOFs can survive the ordering temperature of the spin state. In addition, effective spin Hamiltonians used to describe the phase diagrams of TOFs in an external magnetic field require an amendment by the orbital Zeeman energy.  The latter interaction of TOM with external magnetic fields can be also utilized to control the chirality of the spin texture owing to the close correlation 
between the spin structure and TOM: indeed, interchanging the green and blue atoms in Fig.~\ref{fig:UML_3Q} reverses the sign of the emergent field and the orbital moment, but does not change the energy of the $3Q$ state.

To summarize, we explored the relevance of the modern theory for OM in a set of representatives of diverse
classes of materials, currently under scrutiny in spintronics. Our main message is that outside of the realm of elemental bulk ferromagnets, employing the modern theory description is crucial for understanding orbital magnetism in noncollinear, topologically nontrivial, as well as structurally and chemically heterogeneous systems.

We thank M. dos Santos Dias and S. Heinze for many fruitful discussions. We gratefully acknowledge computing time on the supercomputers JUQUEEN and JURECA at
J\"ulich Supercomputing Center as well as at the JARA-HPC cluster of RWTH Aachen, and funding under SPP 1538 of Deutsche Forschungsgemeinschaft (DFG). H. Z. thanks the LOEWE project RESPONSE funded by the Ministry of Higher Education, Research and the Arts (HMWK) of the Hessen state. A. N. acknowledges financial support from the MAGicSky Horizon 2020 European Research FET Open Project No. 665095.

\bibliography{my_bibliography}

\begin{thebibliography}{55}%
\makeatletter
\providecommand \@ifxundefined [1]{%
 \@ifx{#1\undefined}
}%
\providecommand \@ifnum [1]{%
 \ifnum #1\expandafter \@firstoftwo
 \else \expandafter \@secondoftwo
 \fi
}%
\providecommand \@ifx [1]{%
 \ifx #1\expandafter \@firstoftwo
 \else \expandafter \@secondoftwo
 \fi
}%
\providecommand \natexlab [1]{#1}%
\providecommand \enquote  [1]{``#1''}%
\providecommand \bibnamefont  [1]{#1}%
\providecommand \bibfnamefont [1]{#1}%
\providecommand \citenamefont [1]{#1}%
\providecommand \href@noop [0]{\@secondoftwo}%
\providecommand \href [0]{\begingroup \@sanitize@url \@href}%
\providecommand \@href[1]{\@@startlink{#1}\@@href}%
\providecommand \@@href[1]{\endgroup#1\@@endlink}%
\providecommand \@sanitize@url [0]{\catcode `\\12\catcode `\$12\catcode
  `\&12\catcode `\#12\catcode `\^12\catcode `\_12\catcode `\%12\relax}%
\providecommand \@@startlink[1]{}%
\providecommand \@@endlink[0]{}%
\providecommand \url  [0]{\begingroup\@sanitize@url \@url }%
\providecommand \@url [1]{\endgroup\@href {#1}{\urlprefix }}%
\providecommand \urlprefix  [0]{URL }%
\providecommand \Eprint [0]{\href }%
\providecommand \doibase [0]{http://dx.doi.org/}%
\providecommand \selectlanguage [0]{\@gobble}%
\providecommand \bibinfo  [0]{\@secondoftwo}%
\providecommand \bibfield  [0]{\@secondoftwo}%
\providecommand \translation [1]{[#1]}%
\providecommand \BibitemOpen [0]{}%
\providecommand \bibitemStop [0]{}%
\providecommand \bibitemNoStop [0]{.\EOS\space}%
\providecommand \EOS [0]{\spacefactor3000\relax}%
\providecommand \BibitemShut  [1]{\csname bibitem#1\endcsname}%
\let\auto@bib@innerbib\@empty
\bibitem [{\citenamefont {Meyer}\ and\ \citenamefont {Asch}(1961)}]{Meyer1961}%
  \BibitemOpen
  \bibfield  {author} {\bibinfo {author} {\bibfnamefont {A.}~\bibnamefont
  {Meyer}}\ and\ \bibinfo {author} {\bibfnamefont {G.}~\bibnamefont {Asch}},\
  }\href@noop {} {\bibfield  {journal} {\bibinfo  {journal} {J. Appl. Phys.}\
  }\textbf {\bibinfo {volume} {32}},\ \bibinfo {pages} {330} (\bibinfo {year}
  {1961})}\BibitemShut {NoStop}%
\bibitem [{\citenamefont {Thole}\ \emph {et~al.}(1992)\citenamefont {Thole},
  \citenamefont {Carra}, \citenamefont {Sette},\ and\ \citenamefont {van~der
  Laan}}]{Thole1992}%
  \BibitemOpen
  \bibfield  {author} {\bibinfo {author} {\bibfnamefont {B.~T.}\ \bibnamefont
  {Thole}}, \bibinfo {author} {\bibfnamefont {P.}~\bibnamefont {Carra}},
  \bibinfo {author} {\bibfnamefont {F.}~\bibnamefont {Sette}},\ and\ \bibinfo
  {author} {\bibfnamefont {G.}~\bibnamefont {van~der Laan}},\ }\href@noop {}
  {\bibfield  {journal} {\bibinfo  {journal} {Phys. Rev. Lett.}\ }\textbf
  {\bibinfo {volume} {68}},\ \bibinfo {pages} {1943} (\bibinfo {year}
  {1992})}\BibitemShut {NoStop}%
\bibitem [{\citenamefont {Carra}\ \emph {et~al.}(1993)\citenamefont {Carra},
  \citenamefont {Thole}, \citenamefont {Altarelli},\ and\ \citenamefont
  {Wang}}]{Carra1993}%
  \BibitemOpen
  \bibfield  {author} {\bibinfo {author} {\bibfnamefont {P.}~\bibnamefont
  {Carra}}, \bibinfo {author} {\bibfnamefont {B.~T.}\ \bibnamefont {Thole}},
  \bibinfo {author} {\bibfnamefont {M.}~\bibnamefont {Altarelli}},\ and\
  \bibinfo {author} {\bibfnamefont {X.}~\bibnamefont {Wang}},\ }\href@noop {}
  {\bibfield  {journal} {\bibinfo  {journal} {Phys. Rev. Lett.}\ }\textbf
  {\bibinfo {volume} {70}},\ \bibinfo {pages} {694} (\bibinfo {year}
  {1993})}\BibitemShut {NoStop}%
\bibitem [{\citenamefont {Chen}\ \emph {et~al.}(1995)\citenamefont {Chen},
  \citenamefont {Idzerda}, \citenamefont {Lin}, \citenamefont {Smith},
  \citenamefont {Meigs}, \citenamefont {Chaban}, \citenamefont {Ho},
  \citenamefont {Pellegrin},\ and\ \citenamefont {Sette}}]{Chen1995}%
  \BibitemOpen
  \bibfield  {author} {\bibinfo {author} {\bibfnamefont {C.~T.}\ \bibnamefont
  {Chen}}, \bibinfo {author} {\bibfnamefont {Y.~U.}\ \bibnamefont {Idzerda}},
  \bibinfo {author} {\bibfnamefont {H.-J.}\ \bibnamefont {Lin}}, \bibinfo
  {author} {\bibfnamefont {N.~V.}\ \bibnamefont {Smith}}, \bibinfo {author}
  {\bibfnamefont {G.}~\bibnamefont {Meigs}}, \bibinfo {author} {\bibfnamefont
  {E.}~\bibnamefont {Chaban}}, \bibinfo {author} {\bibfnamefont {G.~H.}\
  \bibnamefont {Ho}}, \bibinfo {author} {\bibfnamefont {E.}~\bibnamefont
  {Pellegrin}},\ and\ \bibinfo {author} {\bibfnamefont {F.}~\bibnamefont
  {Sette}},\ }\href@noop {} {\bibfield  {journal} {\bibinfo  {journal} {Phys.
  Rev. Lett.}\ }\textbf {\bibinfo {volume} {75}},\ \bibinfo {pages} {152}
  (\bibinfo {year} {1995})}\BibitemShut {NoStop}%
\bibitem [{\citenamefont {Taylor}\ \emph {et~al.}(2002)\citenamefont {Taylor},
  \citenamefont {Duffy}, \citenamefont {Bebb}, \citenamefont {Lees},
  \citenamefont {Bouchenoire}, \citenamefont {Brown},\ and\ \citenamefont
  {Cooper}}]{Taylor2002}%
  \BibitemOpen
  \bibfield  {author} {\bibinfo {author} {\bibfnamefont {J.~W.}\ \bibnamefont
  {Taylor}}, \bibinfo {author} {\bibfnamefont {J.~A.}\ \bibnamefont {Duffy}},
  \bibinfo {author} {\bibfnamefont {A.~M.}\ \bibnamefont {Bebb}}, \bibinfo
  {author} {\bibfnamefont {M.~R.}\ \bibnamefont {Lees}}, \bibinfo {author}
  {\bibfnamefont {L.}~\bibnamefont {Bouchenoire}}, \bibinfo {author}
  {\bibfnamefont {S.~D.}\ \bibnamefont {Brown}},\ and\ \bibinfo {author}
  {\bibfnamefont {M.~J.}\ \bibnamefont {Cooper}},\ }\href@noop {} {\bibfield
  {journal} {\bibinfo  {journal} {Phys. Rev. B}\ }\textbf {\bibinfo {volume}
  {66}},\ \bibinfo {pages} {161319(R)} (\bibinfo {year} {2002})}\BibitemShut
  {NoStop}%
\bibitem [{\citenamefont {Gotsis}\ and\ \citenamefont
  {Mazin}(2003)}]{Gotsis2003}%
  \BibitemOpen
  \bibfield  {author} {\bibinfo {author} {\bibfnamefont {H.~J.}\ \bibnamefont
  {Gotsis}}\ and\ \bibinfo {author} {\bibfnamefont {I.~I.}\ \bibnamefont
  {Mazin}},\ }\href@noop {} {\bibfield  {journal} {\bibinfo  {journal} {Phys.
  Rev. B}\ }\textbf {\bibinfo {volume} {68}},\ \bibinfo {pages} {224427}
  (\bibinfo {year} {2003})}\BibitemShut {NoStop}%
\bibitem [{\citenamefont {Qiao}\ \emph {et~al.}(2004)\citenamefont {Qiao},
  \citenamefont {Kimura}, \citenamefont {Adachi}, \citenamefont {Iori},
  \citenamefont {Miyamoto}, \citenamefont {Xie}, \citenamefont {Namatame},
  \citenamefont {Taniguchi}, \citenamefont {Tanaka}, \citenamefont {Muro},
  \citenamefont {Imada},\ and\ \citenamefont {Suga}}]{Qiao2004}%
  \BibitemOpen
  \bibfield  {author} {\bibinfo {author} {\bibfnamefont {S.}~\bibnamefont
  {Qiao}}, \bibinfo {author} {\bibfnamefont {A.}~\bibnamefont {Kimura}},
  \bibinfo {author} {\bibfnamefont {H.}~\bibnamefont {Adachi}}, \bibinfo
  {author} {\bibfnamefont {K.}~\bibnamefont {Iori}}, \bibinfo {author}
  {\bibfnamefont {K.}~\bibnamefont {Miyamoto}}, \bibinfo {author}
  {\bibfnamefont {T.}~\bibnamefont {Xie}}, \bibinfo {author} {\bibfnamefont
  {H.}~\bibnamefont {Namatame}}, \bibinfo {author} {\bibfnamefont
  {M.}~\bibnamefont {Taniguchi}}, \bibinfo {author} {\bibfnamefont
  {A.}~\bibnamefont {Tanaka}}, \bibinfo {author} {\bibfnamefont
  {T.}~\bibnamefont {Muro}}, \bibinfo {author} {\bibfnamefont {S.}~\bibnamefont
  {Imada}},\ and\ \bibinfo {author} {\bibfnamefont {S.}~\bibnamefont {Suga}},\
  }\href@noop {} {\bibfield  {journal} {\bibinfo  {journal} {Phys. Rev. B}\
  }\textbf {\bibinfo {volume} {70}},\ \bibinfo {pages} {134418} (\bibinfo
  {year} {2004})}\BibitemShut {NoStop}%
\bibitem [{\citenamefont {Xiao}\ \emph {et~al.}(2006)\citenamefont {Xiao},
  \citenamefont {Yao}, \citenamefont {Fang},\ and\ \citenamefont
  {Niu}}]{Xiao2006}%
  \BibitemOpen
  \bibfield  {author} {\bibinfo {author} {\bibfnamefont {D.}~\bibnamefont
  {Xiao}}, \bibinfo {author} {\bibfnamefont {Y.}~\bibnamefont {Yao}}, \bibinfo
  {author} {\bibfnamefont {Z.}~\bibnamefont {Fang}},\ and\ \bibinfo {author}
  {\bibfnamefont {Q.}~\bibnamefont {Niu}},\ }\href@noop {} {\bibfield
  {journal} {\bibinfo  {journal} {Phys. Rev. Lett.}\ }\textbf {\bibinfo
  {volume} {97}},\ \bibinfo {pages} {026603} (\bibinfo {year}
  {2006})}\BibitemShut {NoStop}%
\bibitem [{\citenamefont {Murakami}(2006)}]{Murakami2006}%
  \BibitemOpen
  \bibfield  {author} {\bibinfo {author} {\bibfnamefont {S.}~\bibnamefont
  {Murakami}},\ }\href@noop {} {\bibfield  {journal} {\bibinfo  {journal}
  {Phys. Rev. Lett.}\ }\textbf {\bibinfo {volume} {97}},\ \bibinfo {pages}
  {236805} (\bibinfo {year} {2006})}\BibitemShut {NoStop}%
\bibitem [{\citenamefont {Xiao}\ \emph {et~al.}(2007)\citenamefont {Xiao},
  \citenamefont {Yao},\ and\ \citenamefont {Niu}}]{Xiao2007}%
  \BibitemOpen
  \bibfield  {author} {\bibinfo {author} {\bibfnamefont {D.}~\bibnamefont
  {Xiao}}, \bibinfo {author} {\bibfnamefont {W.}~\bibnamefont {Yao}},\ and\
  \bibinfo {author} {\bibfnamefont {Q.}~\bibnamefont {Niu}},\ }\href@noop {}
  {\bibfield  {journal} {\bibinfo  {journal} {Phys. Rev. Lett.}\ }\textbf
  {\bibinfo {volume} {99}},\ \bibinfo {pages} {236809} (\bibinfo {year}
  {2007})}\BibitemShut {NoStop}%
\bibitem [{\citenamefont {Wang}\ and\ \citenamefont {Zhang}(2007)}]{Wang2007}%
  \BibitemOpen
  \bibfield  {author} {\bibinfo {author} {\bibfnamefont {Z.}~\bibnamefont
  {Wang}}\ and\ \bibinfo {author} {\bibfnamefont {P.}~\bibnamefont {Zhang}},\
  }\href@noop {} {\bibfield  {journal} {\bibinfo  {journal} {Phys. Rev. B}\
  }\textbf {\bibinfo {volume} {76}},\ \bibinfo {pages} {064406} (\bibinfo
  {year} {2007})}\BibitemShut {NoStop}%
\bibitem [{\citenamefont {Malashevich}\ \emph {et~al.}(2010)\citenamefont
  {Malashevich}, \citenamefont {Souza}, \citenamefont {Coh},\ and\
  \citenamefont {Vanderbilt}}]{Malashevich2010}%
  \BibitemOpen
  \bibfield  {author} {\bibinfo {author} {\bibfnamefont {A.}~\bibnamefont
  {Malashevich}}, \bibinfo {author} {\bibfnamefont {I.}~\bibnamefont {Souza}},
  \bibinfo {author} {\bibfnamefont {S.}~\bibnamefont {Coh}},\ and\ \bibinfo
  {author} {\bibfnamefont {D.}~\bibnamefont {Vanderbilt}},\ }\href@noop {}
  {\bibfield  {journal} {\bibinfo  {journal} {New J. Phys.}\ }\textbf {\bibinfo
  {volume} {12}},\ \bibinfo {pages} {053032} (\bibinfo {year}
  {2010})}\BibitemShut {NoStop}%
\bibitem [{\citenamefont {Essin}\ \emph {et~al.}(2009)\citenamefont {Essin},
  \citenamefont {Moore},\ and\ \citenamefont {Vanderbilt}}]{Essin2009}%
  \BibitemOpen
  \bibfield  {author} {\bibinfo {author} {\bibfnamefont {A.~M.}\ \bibnamefont
  {Essin}}, \bibinfo {author} {\bibfnamefont {J.~E.}\ \bibnamefont {Moore}},\
  and\ \bibinfo {author} {\bibfnamefont {D.}~\bibnamefont {Vanderbilt}},\
  }\href@noop {} {\bibfield  {journal} {\bibinfo  {journal} {Phys. Rev. Lett.}\
  }\textbf {\bibinfo {volume} {102}},\ \bibinfo {pages} {146805} (\bibinfo
  {year} {2009})}\BibitemShut {NoStop}%
\bibitem [{\citenamefont {Essin}\ \emph {et~al.}(2010)\citenamefont {Essin},
  \citenamefont {Turner}, \citenamefont {Moore},\ and\ \citenamefont
  {Vanderbilt}}]{Essin2010}%
  \BibitemOpen
  \bibfield  {author} {\bibinfo {author} {\bibfnamefont {A.~M.}\ \bibnamefont
  {Essin}}, \bibinfo {author} {\bibfnamefont {A.~M.}\ \bibnamefont {Turner}},
  \bibinfo {author} {\bibfnamefont {J.~E.}\ \bibnamefont {Moore}},\ and\
  \bibinfo {author} {\bibfnamefont {D.}~\bibnamefont {Vanderbilt}},\
  }\href@noop {} {\bibfield  {journal} {\bibinfo  {journal} {Phys. Rev. B}\
  }\textbf {\bibinfo {volume} {81}},\ \bibinfo {pages} {205104} (\bibinfo
  {year} {2010})}\BibitemShut {NoStop}%
\bibitem [{\citenamefont {van Vleck}(1937)}]{Vleck1937}%
  \BibitemOpen
  \bibfield  {author} {\bibinfo {author} {\bibfnamefont {J.~H.}\ \bibnamefont
  {van Vleck}},\ }\href@noop {} {\bibfield  {journal} {\bibinfo  {journal}
  {Phys. Rev.}\ }\textbf {\bibinfo {volume} {52}},\ \bibinfo {pages} {1178}
  (\bibinfo {year} {1937})}\BibitemShut {NoStop}%
\bibitem [{\citenamefont {Moriya}(1960)}]{Moriya1960}%
  \BibitemOpen
  \bibfield  {author} {\bibinfo {author} {\bibfnamefont {T.}~\bibnamefont
  {Moriya}},\ }\href@noop {} {\bibfield  {journal} {\bibinfo  {journal} {Phys.
  Rev.}\ }\textbf {\bibinfo {volume} {120}},\ \bibinfo {pages} {91} (\bibinfo
  {year} {1960})}\BibitemShut {NoStop}%
\bibitem [{\citenamefont {Hoffmann}\ \emph {et~al.}(2015)\citenamefont
  {Hoffmann}, \citenamefont {Weischenberg}, \citenamefont {Dup{\'e}},
  \citenamefont {Freimuth}, \citenamefont {Ferriani}, \citenamefont
  {Mokrousov},\ and\ \citenamefont {Heinze}}]{Hoffmann2015}%
  \BibitemOpen
  \bibfield  {author} {\bibinfo {author} {\bibfnamefont {M.}~\bibnamefont
  {Hoffmann}}, \bibinfo {author} {\bibfnamefont {J.}~\bibnamefont
  {Weischenberg}}, \bibinfo {author} {\bibfnamefont {B.}~\bibnamefont
  {Dup{\'e}}}, \bibinfo {author} {\bibfnamefont {F.}~\bibnamefont {Freimuth}},
  \bibinfo {author} {\bibfnamefont {P.}~\bibnamefont {Ferriani}}, \bibinfo
  {author} {\bibfnamefont {Y.}~\bibnamefont {Mokrousov}},\ and\ \bibinfo
  {author} {\bibfnamefont {S.}~\bibnamefont {Heinze}},\ }\href@noop {}
  {\bibfield  {journal} {\bibinfo  {journal} {Phys. Rev. B}\ }\textbf {\bibinfo
  {volume} {92}},\ \bibinfo {pages} {020401(R)} (\bibinfo {year}
  {2015})}\BibitemShut {NoStop}%
\bibitem [{\citenamefont {Thonhauser}\ \emph {et~al.}(2005)\citenamefont
  {Thonhauser}, \citenamefont {Ceresoli}, \citenamefont {Vanderbilt},\ and\
  \citenamefont {Resta}}]{Thonhauser2005}%
  \BibitemOpen
  \bibfield  {author} {\bibinfo {author} {\bibfnamefont {T.}~\bibnamefont
  {Thonhauser}}, \bibinfo {author} {\bibfnamefont {D.}~\bibnamefont
  {Ceresoli}}, \bibinfo {author} {\bibfnamefont {D.}~\bibnamefont
  {Vanderbilt}},\ and\ \bibinfo {author} {\bibfnamefont {R.}~\bibnamefont
  {Resta}},\ }\href@noop {} {\bibfield  {journal} {\bibinfo  {journal} {Phys.
  Rev. Lett.}\ }\textbf {\bibinfo {volume} {95}},\ \bibinfo {pages} {137205}
  (\bibinfo {year} {2005})}\BibitemShut {NoStop}%
\bibitem [{\citenamefont {Ceresoli}\ \emph {et~al.}(2006)\citenamefont
  {Ceresoli}, \citenamefont {Thonhauser}, \citenamefont {Vanderbilt},\ and\
  \citenamefont {Resta}}]{Ceresoli2006}%
  \BibitemOpen
  \bibfield  {author} {\bibinfo {author} {\bibfnamefont {D.}~\bibnamefont
  {Ceresoli}}, \bibinfo {author} {\bibfnamefont {T.}~\bibnamefont
  {Thonhauser}}, \bibinfo {author} {\bibfnamefont {D.}~\bibnamefont
  {Vanderbilt}},\ and\ \bibinfo {author} {\bibfnamefont {R.}~\bibnamefont
  {Resta}},\ }\href@noop {} {\bibfield  {journal} {\bibinfo  {journal} {Phys.
  Rev. B}\ }\textbf {\bibinfo {volume} {74}},\ \bibinfo {pages} {024408}
  (\bibinfo {year} {2006})}\BibitemShut {NoStop}%
\bibitem [{\citenamefont {Xiao}\ \emph {et~al.}(2005)\citenamefont {Xiao},
  \citenamefont {Shi},\ and\ \citenamefont {Niu}}]{Xiao2005}%
  \BibitemOpen
  \bibfield  {author} {\bibinfo {author} {\bibfnamefont {D.}~\bibnamefont
  {Xiao}}, \bibinfo {author} {\bibfnamefont {J.}~\bibnamefont {Shi}},\ and\
  \bibinfo {author} {\bibfnamefont {Q.}~\bibnamefont {Niu}},\ }\href@noop {}
  {\bibfield  {journal} {\bibinfo  {journal} {Phys. Rev. Lett.}\ }\textbf
  {\bibinfo {volume} {95}},\ \bibinfo {pages} {137204} (\bibinfo {year}
  {2005})}\BibitemShut {NoStop}%
\bibitem [{\citenamefont {Shi}\ \emph {et~al.}(2007)\citenamefont {Shi},
  \citenamefont {Vignale}, \citenamefont {Xiao},\ and\ \citenamefont
  {Niu}}]{Shi2007}%
  \BibitemOpen
  \bibfield  {author} {\bibinfo {author} {\bibfnamefont {J.}~\bibnamefont
  {Shi}}, \bibinfo {author} {\bibfnamefont {G.}~\bibnamefont {Vignale}},
  \bibinfo {author} {\bibfnamefont {D.}~\bibnamefont {Xiao}},\ and\ \bibinfo
  {author} {\bibfnamefont {Q.}~\bibnamefont {Niu}},\ }\href@noop {} {\bibfield
  {journal} {\bibinfo  {journal} {Phys. Rev. Lett.}\ }\textbf {\bibinfo
  {volume} {99}},\ \bibinfo {pages} {197202} (\bibinfo {year}
  {2007})}\BibitemShut {NoStop}%
\bibitem [{\citenamefont {Thonhauser}(2011)}]{Thonhauser2011}%
  \BibitemOpen
  \bibfield  {author} {\bibinfo {author} {\bibfnamefont {T.}~\bibnamefont
  {Thonhauser}},\ }\href@noop {} {\bibfield  {journal} {\bibinfo  {journal}
  {Int. J. Mod. Phys. B}\ }\textbf {\bibinfo {volume} {25}},\ \bibinfo {pages}
  {1429} (\bibinfo {year} {2011})}\BibitemShut {NoStop}%
\bibitem [{\citenamefont {Resta}(2010)}]{Resta2010}%
  \BibitemOpen
  \bibfield  {author} {\bibinfo {author} {\bibfnamefont {R.}~\bibnamefont
  {Resta}},\ }\href@noop {} {\bibfield  {journal} {\bibinfo  {journal} {J.
  Phys.: Condens. Matter}\ }\textbf {\bibinfo {volume} {22}},\ \bibinfo {pages}
  {123201} (\bibinfo {year} {2010})}\BibitemShut {NoStop}%
\bibitem [{\citenamefont {Xiao}\ \emph {et~al.}(2010)\citenamefont {Xiao},
  \citenamefont {Chang},\ and\ \citenamefont {Niu}}]{Xiao2010}%
  \BibitemOpen
  \bibfield  {author} {\bibinfo {author} {\bibfnamefont {D.}~\bibnamefont
  {Xiao}}, \bibinfo {author} {\bibfnamefont {M.-C.}\ \bibnamefont {Chang}},\
  and\ \bibinfo {author} {\bibfnamefont {Q.}~\bibnamefont {Niu}},\ }\href@noop
  {} {\bibfield  {journal} {\bibinfo  {journal} {Rev. Mod. Phys.}\ }\textbf
  {\bibinfo {volume} {82}},\ \bibinfo {pages} {1959} (\bibinfo {year}
  {2010})}\BibitemShut {NoStop}%
\bibitem [{\citenamefont {Nikolaev}\ and\ \citenamefont
  {Solovyev}(2014)}]{Nikolaev2014}%
  \BibitemOpen
  \bibfield  {author} {\bibinfo {author} {\bibfnamefont {S.~A.}\ \bibnamefont
  {Nikolaev}}\ and\ \bibinfo {author} {\bibfnamefont {I.~V.}\ \bibnamefont
  {Solovyev}},\ }\href@noop {} {\bibfield  {journal} {\bibinfo  {journal}
  {Phys. Rev. B}\ }\textbf {\bibinfo {volume} {89}},\ \bibinfo {pages} {064428}
  (\bibinfo {year} {2014})}\BibitemShut {NoStop}%
\bibitem [{\citenamefont {Miron}\ \emph {et~al.}(2011)\citenamefont {Miron},
  \citenamefont {Garello}, \citenamefont {Gaudin}, \citenamefont {Zermatten},
  \citenamefont {Costache}, \citenamefont {Auffret}, \citenamefont {Bandiera},
  \citenamefont {Rodmacq}, \citenamefont {Schuhl},\ and\ \citenamefont
  {Gambardella}}]{Miron2011}%
  \BibitemOpen
  \bibfield  {author} {\bibinfo {author} {\bibfnamefont {I.~M.}\ \bibnamefont
  {Miron}}, \bibinfo {author} {\bibfnamefont {K.}~\bibnamefont {Garello}},
  \bibinfo {author} {\bibfnamefont {G.}~\bibnamefont {Gaudin}}, \bibinfo
  {author} {\bibfnamefont {P.-J.}\ \bibnamefont {Zermatten}}, \bibinfo {author}
  {\bibfnamefont {M.~V.}\ \bibnamefont {Costache}}, \bibinfo {author}
  {\bibfnamefont {S.}~\bibnamefont {Auffret}}, \bibinfo {author} {\bibfnamefont
  {S.}~\bibnamefont {Bandiera}}, \bibinfo {author} {\bibfnamefont
  {B.}~\bibnamefont {Rodmacq}}, \bibinfo {author} {\bibfnamefont
  {A.}~\bibnamefont {Schuhl}},\ and\ \bibinfo {author} {\bibfnamefont
  {P.}~\bibnamefont {Gambardella}},\ }\href@noop {} {\bibfield  {journal}
  {\bibinfo  {journal} {Nature}\ }\textbf {\bibinfo {volume} {476}},\ \bibinfo
  {pages} {189} (\bibinfo {year} {2011})}\BibitemShut {NoStop}%
\bibitem [{\citenamefont {Chang}\ \emph {et~al.}(2013)\citenamefont {Chang},
  \citenamefont {Zhang}, \citenamefont {Feng}, \citenamefont {Shen},
  \citenamefont {Zhang}, \citenamefont {Guo}, \citenamefont {Li}, \citenamefont
  {Ou}, \citenamefont {Wei}, \citenamefont {Wang}, \citenamefont {Ji},
  \citenamefont {Feng}, \citenamefont {Ji}, \citenamefont {Chen}, \citenamefont
  {Jia}, \citenamefont {Dai}, \citenamefont {Fang}, \citenamefont {Zhang},
  \citenamefont {He}, \citenamefont {Wang}, \citenamefont {Lu}, \citenamefont
  {Ma},\ and\ \citenamefont {Xue}}]{Chang2013}%
  \BibitemOpen
  \bibfield  {author} {\bibinfo {author} {\bibfnamefont {C.-Z.}\ \bibnamefont
  {Chang}}, \bibinfo {author} {\bibfnamefont {J.}~\bibnamefont {Zhang}},
  \bibinfo {author} {\bibfnamefont {X.}~\bibnamefont {Feng}}, \bibinfo {author}
  {\bibfnamefont {J.}~\bibnamefont {Shen}}, \bibinfo {author} {\bibfnamefont
  {Z.}~\bibnamefont {Zhang}}, \bibinfo {author} {\bibfnamefont
  {M.}~\bibnamefont {Guo}}, \bibinfo {author} {\bibfnamefont {K.}~\bibnamefont
  {Li}}, \bibinfo {author} {\bibfnamefont {Y.}~\bibnamefont {Ou}}, \bibinfo
  {author} {\bibfnamefont {P.}~\bibnamefont {Wei}}, \bibinfo {author}
  {\bibfnamefont {L.-L.}\ \bibnamefont {Wang}}, \bibinfo {author}
  {\bibfnamefont {Z.-Q.}\ \bibnamefont {Ji}}, \bibinfo {author} {\bibfnamefont
  {Y.}~\bibnamefont {Feng}}, \bibinfo {author} {\bibfnamefont {S.}~\bibnamefont
  {Ji}}, \bibinfo {author} {\bibfnamefont {X.}~\bibnamefont {Chen}}, \bibinfo
  {author} {\bibfnamefont {J.}~\bibnamefont {Jia}}, \bibinfo {author}
  {\bibfnamefont {X.}~\bibnamefont {Dai}}, \bibinfo {author} {\bibfnamefont
  {Z.}~\bibnamefont {Fang}}, \bibinfo {author} {\bibfnamefont {S.-C.}\
  \bibnamefont {Zhang}}, \bibinfo {author} {\bibfnamefont {K.}~\bibnamefont
  {He}}, \bibinfo {author} {\bibfnamefont {Y.}~\bibnamefont {Wang}}, \bibinfo
  {author} {\bibfnamefont {L.}~\bibnamefont {Lu}}, \bibinfo {author}
  {\bibfnamefont {X.-C.}\ \bibnamefont {Ma}},\ and\ \bibinfo {author}
  {\bibfnamefont {Q.-K.}\ \bibnamefont {Xue}},\ }\href@noop {} {\bibfield
  {journal} {\bibinfo  {journal} {Science}\ }\textbf {\bibinfo {volume}
  {340}},\ \bibinfo {pages} {167} (\bibinfo {year} {2013})}\BibitemShut
  {NoStop}%
\bibitem [{\citenamefont {Taguchi}\ \emph {et~al.}(2001)\citenamefont
  {Taguchi}, \citenamefont {Oohara}, \citenamefont {Yoshizawa}, \citenamefont
  {Nagaosa},\ and\ \citenamefont {Tokura}}]{Taguchi2001}%
  \BibitemOpen
  \bibfield  {author} {\bibinfo {author} {\bibfnamefont {Y.}~\bibnamefont
  {Taguchi}}, \bibinfo {author} {\bibfnamefont {Y.}~\bibnamefont {Oohara}},
  \bibinfo {author} {\bibfnamefont {H.}~\bibnamefont {Yoshizawa}}, \bibinfo
  {author} {\bibfnamefont {N.}~\bibnamefont {Nagaosa}},\ and\ \bibinfo
  {author} {\bibfnamefont {Y.}~\bibnamefont {Tokura}},\ }\href@noop {}
  {\bibfield  {journal} {\bibinfo  {journal} {Science}\ }\textbf {\bibinfo
  {volume} {291}},\ \bibinfo {pages} {2573} (\bibinfo {year}
  {2001})}\BibitemShut {NoStop}%
\bibitem [{\citenamefont {Heinze}\ \emph {et~al.}(2011)\citenamefont {Heinze},
  \citenamefont {Von~Bergmann}, \citenamefont {Menzel}, \citenamefont {Brede},
  \citenamefont {Kubetzka}, \citenamefont {Wiesendanger}, \citenamefont
  {Bihlmayer},\ and\ \citenamefont {Bl{\"u}gel}}]{Heinze2011}%
  \BibitemOpen
  \bibfield  {author} {\bibinfo {author} {\bibfnamefont {S.}~\bibnamefont
  {Heinze}}, \bibinfo {author} {\bibfnamefont {K.}~\bibnamefont
  {Von~Bergmann}}, \bibinfo {author} {\bibfnamefont {M.}~\bibnamefont
  {Menzel}}, \bibinfo {author} {\bibfnamefont {J.}~\bibnamefont {Brede}},
  \bibinfo {author} {\bibfnamefont {A.}~\bibnamefont {Kubetzka}}, \bibinfo
  {author} {\bibfnamefont {R.}~\bibnamefont {Wiesendanger}}, \bibinfo {author}
  {\bibfnamefont {G.}~\bibnamefont {Bihlmayer}},\ and\ \bibinfo {author}
  {\bibfnamefont {S.}~\bibnamefont {Bl{\"u}gel}},\ }\href@noop {} {\bibfield
  {journal} {\bibinfo  {journal} {Nat. Phys.}\ }\textbf {\bibinfo {volume}
  {7}},\ \bibinfo {pages} {713} (\bibinfo {year} {2011})}\BibitemShut {NoStop}%
\bibitem [{\citenamefont {Ceresoli}\ \emph {et~al.}(2010)\citenamefont
  {Ceresoli}, \citenamefont {Gerstmann}, \citenamefont {Seitsonen},\ and\
  \citenamefont {Mauri}}]{Ceresoli2010}%
  \BibitemOpen
  \bibfield  {author} {\bibinfo {author} {\bibfnamefont {D.}~\bibnamefont
  {Ceresoli}}, \bibinfo {author} {\bibfnamefont {U.}~\bibnamefont {Gerstmann}},
  \bibinfo {author} {\bibfnamefont {A.~P.}\ \bibnamefont {Seitsonen}},\ and\
  \bibinfo {author} {\bibfnamefont {F.}~\bibnamefont {Mauri}},\ }\href@noop {}
  {\bibfield  {journal} {\bibinfo  {journal} {Phys. Rev. B}\ }\textbf {\bibinfo
  {volume} {81}},\ \bibinfo {pages} {060409(R)} (\bibinfo {year}
  {2010})}\BibitemShut {NoStop}%
\bibitem [{\citenamefont {Lopez}\ \emph {et~al.}(2012)\citenamefont {Lopez},
  \citenamefont {Vanderbilt}, \citenamefont {Thonhauser},\ and\ \citenamefont
  {Souza}}]{Lopez2012}%
  \BibitemOpen
  \bibfield  {author} {\bibinfo {author} {\bibfnamefont {M.~G.}\ \bibnamefont
  {Lopez}}, \bibinfo {author} {\bibfnamefont {D.}~\bibnamefont {Vanderbilt}},
  \bibinfo {author} {\bibfnamefont {T.}~\bibnamefont {Thonhauser}},\ and\
  \bibinfo {author} {\bibfnamefont {I.}~\bibnamefont {Souza}},\ }\href@noop {}
  {\bibfield  {journal} {\bibinfo  {journal} {Phys. Rev. B}\ }\textbf {\bibinfo
  {volume} {85}},\ \bibinfo {pages} {014435} (\bibinfo {year}
  {2012})}\BibitemShut {NoStop}%
\bibitem [{\citenamefont {Yates}\ \emph {et~al.}(2007)\citenamefont {Yates},
  \citenamefont {Wang}, \citenamefont {Vanderbilt},\ and\ \citenamefont
  {Souza}}]{Yates2007}%
  \BibitemOpen
  \bibfield  {author} {\bibinfo {author} {\bibfnamefont {J.~R.}\ \bibnamefont
  {Yates}}, \bibinfo {author} {\bibfnamefont {X.}~\bibnamefont {Wang}},
  \bibinfo {author} {\bibfnamefont {D.}~\bibnamefont {Vanderbilt}},\ and\
  \bibinfo {author} {\bibfnamefont {I.}~\bibnamefont {Souza}},\ }\href@noop {}
  {\bibfield  {journal} {\bibinfo  {journal} {Phys. Rev. B}\ }\textbf {\bibinfo
  {volume} {75}},\ \bibinfo {pages} {195121} (\bibinfo {year}
  {2007})}\BibitemShut {NoStop}%
\bibitem [{sup()}]{suppl}%
  \BibitemOpen
  \href@noop {} {}\bibinfo {howpublished} {See Supplemental Material for details on first-principles calculation and Wannier
  interpolation}\BibitemShut {NoStop}%
\bibitem [{fle()}]{fleur}%
  \BibitemOpen
  \href@noop {} {}\bibinfo {howpublished} {See
  \url{http://www.flapw.de}}\BibitemShut {NoStop}%
\bibitem [{\citenamefont {Marzari}\ and\ \citenamefont
  {Vanderbilt}(1997)}]{Marzari1997}%
  \BibitemOpen
  \bibfield  {author} {\bibinfo {author} {\bibfnamefont {N.}~\bibnamefont
  {Marzari}}\ and\ \bibinfo {author} {\bibfnamefont {D.}~\bibnamefont
  {Vanderbilt}},\ }\href@noop {} {\bibfield  {journal} {\bibinfo  {journal}
  {Phys. Rev. B}\ }\textbf {\bibinfo {volume} {56}},\ \bibinfo {pages} {12847}
  (\bibinfo {year} {1997})}\BibitemShut {NoStop}%
\bibitem [{\citenamefont {Souza}\ \emph {et~al.}(2001)\citenamefont {Souza},
  \citenamefont {Marzari},\ and\ \citenamefont {Vanderbilt}}]{Souza2001}%
  \BibitemOpen
  \bibfield  {author} {\bibinfo {author} {\bibfnamefont {I.}~\bibnamefont
  {Souza}}, \bibinfo {author} {\bibfnamefont {N.}~\bibnamefont {Marzari}},\
  and\ \bibinfo {author} {\bibfnamefont {D.}~\bibnamefont {Vanderbilt}},\
  }\href@noop {} {\bibfield  {journal} {\bibinfo  {journal} {Phys. Rev. B}\
  }\textbf {\bibinfo {volume} {65}},\ \bibinfo {pages} {035109} (\bibinfo
  {year} {2001})}\BibitemShut {NoStop}%
\bibitem [{\citenamefont {Marzari}\ \emph {et~al.}(2012)\citenamefont
  {Marzari}, \citenamefont {Mostofi}, \citenamefont {Yates}, \citenamefont
  {Souza},\ and\ \citenamefont {Vanderbilt}}]{Marzari2012}%
  \BibitemOpen
  \bibfield  {author} {\bibinfo {author} {\bibfnamefont {N.}~\bibnamefont
  {Marzari}}, \bibinfo {author} {\bibfnamefont {A.~A.}\ \bibnamefont
  {Mostofi}}, \bibinfo {author} {\bibfnamefont {J.~R.}\ \bibnamefont {Yates}},
  \bibinfo {author} {\bibfnamefont {I.}~\bibnamefont {Souza}},\ and\ \bibinfo
  {author} {\bibfnamefont {D.}~\bibnamefont {Vanderbilt}},\ }\href@noop {}
  {\bibfield  {journal} {\bibinfo  {journal} {Rev. Mod. Phys.}\ }\textbf
  {\bibinfo {volume} {84}},\ \bibinfo {pages} {1419} (\bibinfo {year}
  {2012})}\BibitemShut {NoStop}%
\bibitem [{\citenamefont {Mostofi}\ \emph {et~al.}(2008)\citenamefont
  {Mostofi}, \citenamefont {Yates}, \citenamefont {Lee}, \citenamefont {Souza},
  \citenamefont {Vanderbilt},\ and\ \citenamefont {Marzari}}]{Mostofi2008}%
  \BibitemOpen
  \bibfield  {author} {\bibinfo {author} {\bibfnamefont {A.~A.}\ \bibnamefont
  {Mostofi}}, \bibinfo {author} {\bibfnamefont {J.~R.}\ \bibnamefont {Yates}},
  \bibinfo {author} {\bibfnamefont {Y.-S.}\ \bibnamefont {Lee}}, \bibinfo
  {author} {\bibfnamefont {I.}~\bibnamefont {Souza}}, \bibinfo {author}
  {\bibfnamefont {D.}~\bibnamefont {Vanderbilt}},\ and\ \bibinfo {author}
  {\bibfnamefont {N.}~\bibnamefont {Marzari}},\ }\href@noop {} {\bibfield
  {journal} {\bibinfo  {journal} {Comp. Phys. Commun.}\ }\textbf {\bibinfo
  {volume} {178}},\ \bibinfo {pages} {685} (\bibinfo {year}
  {2008})}\BibitemShut {NoStop}%
\bibitem [{\citenamefont {Freimuth}\ \emph {et~al.}(2008)\citenamefont
  {Freimuth}, \citenamefont {Mokrousov}, \citenamefont {Wortmann},
  \citenamefont {Heinze},\ and\ \citenamefont {Bl{\"u}gel}}]{Freimuth2008}%
  \BibitemOpen
  \bibfield  {author} {\bibinfo {author} {\bibfnamefont {F.}~\bibnamefont
  {Freimuth}}, \bibinfo {author} {\bibfnamefont {Y.}~\bibnamefont {Mokrousov}},
  \bibinfo {author} {\bibfnamefont {D.}~\bibnamefont {Wortmann}}, \bibinfo
  {author} {\bibfnamefont {S.}~\bibnamefont {Heinze}},\ and\ \bibinfo {author}
  {\bibfnamefont {S.}~\bibnamefont {Bl{\"u}gel}},\ }\href@noop {} {\bibfield
  {journal} {\bibinfo  {journal} {Phys. Rev. B}\ }\textbf {\bibinfo {volume}
  {78}},\ \bibinfo {pages} {035120} (\bibinfo {year} {2008})}\BibitemShut
  {NoStop}%
\bibitem [{\citenamefont {Ferriani}\ \emph {et~al.}(2005)\citenamefont
  {Ferriani}, \citenamefont {Heinze}, \citenamefont {Bihlmayer},\ and\
  \citenamefont {Bl{\"u}gel}}]{Ferriani2005}%
  \BibitemOpen
  \bibfield  {author} {\bibinfo {author} {\bibfnamefont {P.}~\bibnamefont
  {Ferriani}}, \bibinfo {author} {\bibfnamefont {S.}~\bibnamefont {Heinze}},
  \bibinfo {author} {\bibfnamefont {G.}~\bibnamefont {Bihlmayer}},\ and\
  \bibinfo {author} {\bibfnamefont {S.}~\bibnamefont {Bl{\"u}gel}},\
  }\href@noop {} {\bibfield  {journal} {\bibinfo  {journal} {Phys. Rev. B}\
  }\textbf {\bibinfo {volume} {72}},\ \bibinfo {pages} {024452} (\bibinfo
  {year} {2005})}\BibitemShut {NoStop}%
\bibitem [{\citenamefont {Ferriani}\ \emph {et~al.}(2008)\citenamefont
  {Ferriani}, \citenamefont {Von~Bergmann}, \citenamefont {Vedmedenko},
  \citenamefont {Heinze}, \citenamefont {Bode}, \citenamefont {Heide},
  \citenamefont {Bihlmayer}, \citenamefont {Bl{\"u}gel},\ and\ \citenamefont
  {Wiesendanger}}]{Ferriani2008}%
  \BibitemOpen
  \bibfield  {author} {\bibinfo {author} {\bibfnamefont {P.}~\bibnamefont
  {Ferriani}}, \bibinfo {author} {\bibfnamefont {K.}~\bibnamefont
  {Von~Bergmann}}, \bibinfo {author} {\bibfnamefont {E.~Y.}\ \bibnamefont
  {Vedmedenko}}, \bibinfo {author} {\bibfnamefont {S.}~\bibnamefont {Heinze}},
  \bibinfo {author} {\bibfnamefont {M.}~\bibnamefont {Bode}}, \bibinfo {author}
  {\bibfnamefont {M.}~\bibnamefont {Heide}}, \bibinfo {author} {\bibfnamefont
  {G.}~\bibnamefont {Bihlmayer}}, \bibinfo {author} {\bibfnamefont
  {S.}~\bibnamefont {Bl{\"u}gel}},\ and\ \bibinfo {author} {\bibfnamefont
  {R.}~\bibnamefont {Wiesendanger}},\ }\href@noop {} {\bibfield  {journal}
  {\bibinfo  {journal} {Phys. Rev. Lett.}\ }\textbf {\bibinfo {volume} {101}},\
  \bibinfo {pages} {027201} (\bibinfo {year} {2008})}\BibitemShut {NoStop}%
\bibitem [{\citenamefont {Ferriani}\ \emph {et~al.}(2009)\citenamefont
  {Ferriani}, \citenamefont {Von~Bergmann}, \citenamefont {Vedmedenko},
  \citenamefont {Heinze}, \citenamefont {Bode}, \citenamefont {Heide},
  \citenamefont {Bihlmayer}, \citenamefont {Bl{\"u}gel},\ and\ \citenamefont
  {Wiesendanger}}]{Ferriani2009}%
  \BibitemOpen
  \bibfield  {author} {\bibinfo {author} {\bibfnamefont {P.}~\bibnamefont
  {Ferriani}}, \bibinfo {author} {\bibfnamefont {K.}~\bibnamefont
  {Von~Bergmann}}, \bibinfo {author} {\bibfnamefont {E.~Y.}\ \bibnamefont
  {Vedmedenko}}, \bibinfo {author} {\bibfnamefont {S.}~\bibnamefont {Heinze}},
  \bibinfo {author} {\bibfnamefont {M.}~\bibnamefont {Bode}}, \bibinfo {author}
  {\bibfnamefont {M.}~\bibnamefont {Heide}}, \bibinfo {author} {\bibfnamefont
  {G.}~\bibnamefont {Bihlmayer}}, \bibinfo {author} {\bibfnamefont
  {S.}~\bibnamefont {Bl{\"u}gel}},\ and\ \bibinfo {author} {\bibfnamefont
  {R.}~\bibnamefont {Wiesendanger}},\ }\href@noop {} {\bibfield  {journal}
  {\bibinfo  {journal} {Phys. Rev. Lett.}\ }\textbf {\bibinfo {volume} {102}},\
  \bibinfo {pages} {019901(E)} (\bibinfo {year} {2009})}\BibitemShut {NoStop}%
\bibitem [{\citenamefont {Zhang}\ \emph {et~al.}(2012)\citenamefont {Zhang},
  \citenamefont {Lazo}, \citenamefont {Bl{\"u}gel}, \citenamefont {Heinze},\
  and\ \citenamefont {Mokrousov}}]{Zhang2012}%
  \BibitemOpen
  \bibfield  {author} {\bibinfo {author} {\bibfnamefont {H.}~\bibnamefont
  {Zhang}}, \bibinfo {author} {\bibfnamefont {C.}~\bibnamefont {Lazo}},
  \bibinfo {author} {\bibfnamefont {S.}~\bibnamefont {Bl{\"u}gel}}, \bibinfo
  {author} {\bibfnamefont {S.}~\bibnamefont {Heinze}},\ and\ \bibinfo {author}
  {\bibfnamefont {Y.}~\bibnamefont {Mokrousov}},\ }\href@noop {} {\bibfield
  {journal} {\bibinfo  {journal} {Phys. Rev. Lett.}\ }\textbf {\bibinfo
  {volume} {108}},\ \bibinfo {pages} {056802} (\bibinfo {year}
  {2012})}\BibitemShut {NoStop}%
\bibitem [{Note1()}]{Note1}%
  \BibitemOpen
  \bibinfo {note} {The atomic coordinates and computational parameters from
  Ref.~\cite {Zhang2012} were used}\BibitemShut {NoStop}%
\bibitem [{\citenamefont {Kurz}\ \emph {et~al.}(2001)\citenamefont {Kurz},
  \citenamefont {Bihlmayer}, \citenamefont {Hirai},\ and\ \citenamefont
  {Bl{\"u}gel}}]{Kurz2001}%
  \BibitemOpen
  \bibfield  {author} {\bibinfo {author} {\bibfnamefont {P.}~\bibnamefont
  {Kurz}}, \bibinfo {author} {\bibfnamefont {G.}~\bibnamefont {Bihlmayer}},
  \bibinfo {author} {\bibfnamefont {K.}~\bibnamefont {Hirai}},\ and\ \bibinfo
  {author} {\bibfnamefont {S.}~\bibnamefont {Bl{\"u}gel}},\ }\href@noop {}
  {\bibfield  {journal} {\bibinfo  {journal} {Phys. Rev. Lett.}\ }\textbf
  {\bibinfo {volume} {86}},\ \bibinfo {pages} {1106} (\bibinfo {year}
  {2001})}\BibitemShut {NoStop}%
\bibitem [{\citenamefont {Dzyaloshinsky}(1958)}]{Dzyaloshinsky1958}%
  \BibitemOpen
  \bibfield  {author} {\bibinfo {author} {\bibfnamefont {I.}~\bibnamefont
  {Dzyaloshinsky}},\ }\href@noop {} {\bibfield  {journal} {\bibinfo  {journal}
  {J. Phys. Chem. Solids}\ }\textbf {\bibinfo {volume} {4}},\ \bibinfo {pages}
  {241} (\bibinfo {year} {1958})}\BibitemShut {NoStop}%
\bibitem [{\citenamefont {Franz}\ \emph {et~al.}(2014)\citenamefont {Franz},
  \citenamefont {Freimuth}, \citenamefont {Bauer}, \citenamefont {Ritz},
  \citenamefont {Schnarr}, \citenamefont {Duvinage}, \citenamefont {Adams},
  \citenamefont {Bl{\"u}gel}, \citenamefont {Rosch}, \citenamefont
  {Mokrousov},\ and\ \citenamefont {Pfleiderer}}]{Franz2014}%
  \BibitemOpen
  \bibfield  {author} {\bibinfo {author} {\bibfnamefont {C.}~\bibnamefont
  {Franz}}, \bibinfo {author} {\bibfnamefont {F.}~\bibnamefont {Freimuth}},
  \bibinfo {author} {\bibfnamefont {A.}~\bibnamefont {Bauer}}, \bibinfo
  {author} {\bibfnamefont {R.}~\bibnamefont {Ritz}}, \bibinfo {author}
  {\bibfnamefont {C.}~\bibnamefont {Schnarr}}, \bibinfo {author} {\bibfnamefont
  {C.}~\bibnamefont {Duvinage}}, \bibinfo {author} {\bibfnamefont
  {T.}~\bibnamefont {Adams}}, \bibinfo {author} {\bibfnamefont
  {S.}~\bibnamefont {Bl{\"u}gel}}, \bibinfo {author} {\bibfnamefont
  {A.}~\bibnamefont {Rosch}}, \bibinfo {author} {\bibfnamefont
  {Y.}~\bibnamefont {Mokrousov}},\ and\ \bibinfo {author} {\bibfnamefont
  {C.}~\bibnamefont {Pfleiderer}},\ }\href@noop {} {\bibfield  {journal}
  {\bibinfo  {journal} {Phys. Rev. Lett.}\ }\textbf {\bibinfo {volume} {112}},\
  \bibinfo {pages} {186601} (\bibinfo {year} {2014})}\BibitemShut {NoStop}%
\bibitem [{\citenamefont {K{\"u}bler}\ and\ \citenamefont
  {Felser}(2014)}]{Kuebler2014}%
  \BibitemOpen
  \bibfield  {author} {\bibinfo {author} {\bibfnamefont {J.}~\bibnamefont
  {K{\"u}bler}}\ and\ \bibinfo {author} {\bibfnamefont {C.}~\bibnamefont
  {Felser}},\ }\href@noop {} {\bibfield  {journal} {\bibinfo  {journal}
  {Europhys. Lett.}\ }\textbf {\bibinfo {volume} {108}},\ \bibinfo {pages}
  {67001} (\bibinfo {year} {2014})}\BibitemShut {NoStop}%
\bibitem [{\citenamefont {Shindou}\ and\ \citenamefont
  {Nagaosa}(2001)}]{Shindou2001}%
  \BibitemOpen
  \bibfield  {author} {\bibinfo {author} {\bibfnamefont {R.}~\bibnamefont
  {Shindou}}\ and\ \bibinfo {author} {\bibfnamefont {N.}~\bibnamefont
  {Nagaosa}},\ }\href@noop {} {\bibfield  {journal} {\bibinfo  {journal} {Phys.
  Rev. Lett.}\ }\textbf {\bibinfo {volume} {87}},\ \bibinfo {pages} {116801}
  (\bibinfo {year} {2001})}\BibitemShut {NoStop}%
\bibitem [{\citenamefont {Bruno}\ \emph {et~al.}(2004)\citenamefont {Bruno},
  \citenamefont {Dugaev},\ and\ \citenamefont {Taillefumier}}]{Bruno2004}%
  \BibitemOpen
  \bibfield  {author} {\bibinfo {author} {\bibfnamefont {P.}~\bibnamefont
  {Bruno}}, \bibinfo {author} {\bibfnamefont {V.~K.}\ \bibnamefont {Dugaev}},\
  and\ \bibinfo {author} {\bibfnamefont {M.}~\bibnamefont {Taillefumier}},\
  }\href@noop {} {\bibfield  {journal} {\bibinfo  {journal} {Phys. Rev. Lett.}\
  }\textbf {\bibinfo {volume} {93}},\ \bibinfo {pages} {096806} (\bibinfo
  {year} {2004})}\BibitemShut {NoStop}%
\bibitem [{\citenamefont {Neubauer}\ \emph {et~al.}(2009)\citenamefont
  {Neubauer}, \citenamefont {Pfleiderer}, \citenamefont {Binz}, \citenamefont
  {Rosch}, \citenamefont {Ritz}, \citenamefont {Niklowitz},\ and\ \citenamefont
  {B{\"o}ni}}]{Neubauer2009}%
  \BibitemOpen
  \bibfield  {author} {\bibinfo {author} {\bibfnamefont {A.}~\bibnamefont
  {Neubauer}}, \bibinfo {author} {\bibfnamefont {C.}~\bibnamefont
  {Pfleiderer}}, \bibinfo {author} {\bibfnamefont {B.}~\bibnamefont {Binz}},
  \bibinfo {author} {\bibfnamefont {A.}~\bibnamefont {Rosch}}, \bibinfo
  {author} {\bibfnamefont {R.}~\bibnamefont {Ritz}}, \bibinfo {author}
  {\bibfnamefont {P.~G.}\ \bibnamefont {Niklowitz}},\ and\ \bibinfo {author}
  {\bibfnamefont {P.}~\bibnamefont {B{\"o}ni}},\ }\href@noop {} {\bibfield
  {journal} {\bibinfo  {journal} {Phys. Rev. Lett.}\ }\textbf {\bibinfo
  {volume} {102}},\ \bibinfo {pages} {186602} (\bibinfo {year}
  {2009})}\BibitemShut {NoStop}%
\bibitem [{\citenamefont {Kanazawa}\ \emph {et~al.}(2011)\citenamefont
  {Kanazawa}, \citenamefont {Onose}, \citenamefont {Arima}, \citenamefont
  {Okuyama}, \citenamefont {Ohoyama}, \citenamefont {Wakimoto}, \citenamefont
  {Kakurai}, \citenamefont {Ishiwata},\ and\ \citenamefont
  {Tokura}}]{Kanazawa2011}%
  \BibitemOpen
  \bibfield  {author} {\bibinfo {author} {\bibfnamefont {N.}~\bibnamefont
  {Kanazawa}}, \bibinfo {author} {\bibfnamefont {Y.}~\bibnamefont {Onose}},
  \bibinfo {author} {\bibfnamefont {T.}~\bibnamefont {Arima}}, \bibinfo
  {author} {\bibfnamefont {D.}~\bibnamefont {Okuyama}}, \bibinfo {author}
  {\bibfnamefont {K.}~\bibnamefont {Ohoyama}}, \bibinfo {author} {\bibfnamefont
  {S.}~\bibnamefont {Wakimoto}}, \bibinfo {author} {\bibfnamefont
  {K.}~\bibnamefont {Kakurai}}, \bibinfo {author} {\bibfnamefont
  {S.}~\bibnamefont {Ishiwata}},\ and\ \bibinfo {author} {\bibfnamefont
  {Y.}~\bibnamefont {Tokura}},\ }\href@noop {} {\bibfield  {journal} {\bibinfo
  {journal} {Phys. Rev. Lett.}\ }\textbf {\bibinfo {volume} {106}},\ \bibinfo
  {pages} {156603} (\bibinfo {year} {2011})}\BibitemShut {NoStop}%
\bibitem [{\citenamefont {Huang}\ and\ \citenamefont
  {Chien}(2012)}]{Huang2012}%
  \BibitemOpen
  \bibfield  {author} {\bibinfo {author} {\bibfnamefont {S.~X.}\ \bibnamefont
  {Huang}}\ and\ \bibinfo {author} {\bibfnamefont {C.~L.}\ \bibnamefont
  {Chien}},\ }\href@noop {} {\bibfield  {journal} {\bibinfo  {journal} {Phys.
  Rev. Lett.}\ }\textbf {\bibinfo {volume} {108}},\ \bibinfo {pages} {267201}
  (\bibinfo {year} {2012})}\BibitemShut {NoStop}%
\bibitem [{\citenamefont {Gayles}\ \emph {et~al.}(2015)\citenamefont {Gayles},
  \citenamefont {Freimuth}, \citenamefont {Schena}, \citenamefont {Lani},
  \citenamefont {Mavropoulos}, \citenamefont {Duine}, \citenamefont
  {Bl{\"u}gel}, \citenamefont {Sinova},\ and\ \citenamefont
  {Mokrousov}}]{Gayles2015}%
  \BibitemOpen
  \bibfield  {author} {\bibinfo {author} {\bibfnamefont {J.}~\bibnamefont
  {Gayles}}, \bibinfo {author} {\bibfnamefont {F.}~\bibnamefont {Freimuth}},
  \bibinfo {author} {\bibfnamefont {T.}~\bibnamefont {Schena}}, \bibinfo
  {author} {\bibfnamefont {G.}~\bibnamefont {Lani}}, \bibinfo {author}
  {\bibfnamefont {P.}~\bibnamefont {Mavropoulos}}, \bibinfo {author}
  {\bibfnamefont {R.~A.}\ \bibnamefont {Duine}}, \bibinfo {author}
  {\bibfnamefont {S.}~\bibnamefont {Bl{\"u}gel}}, \bibinfo {author}
  {\bibfnamefont {J.}~\bibnamefont {Sinova}},\ and\ \bibinfo {author}
  {\bibfnamefont {Y.}~\bibnamefont {Mokrousov}},\ }\href@noop {} {\bibfield
  {journal} {\bibinfo  {journal} {Phys. Rev. Lett.}\ }\textbf {\bibinfo
  {volume} {115}},\ \bibinfo {pages} {036602} (\bibinfo {year}
  {2015})}\BibitemShut {NoStop}%
\bibitem [{\citenamefont {Menzel}\ \emph {et~al.}(2012)\citenamefont {Menzel},
  \citenamefont {Mokrousov}, \citenamefont {Wieser}, \citenamefont {Bickel},
  \citenamefont {Vedmedenko}, \citenamefont {Bl{\"u}gel}, \citenamefont
  {Heinze}, \citenamefont {von Bergmann}, \citenamefont {Kubetzka},\ and\
  \citenamefont {Wiesendanger}}]{Menzel2012}%
  \BibitemOpen
  \bibfield  {author} {\bibinfo {author} {\bibfnamefont {M.}~\bibnamefont
  {Menzel}}, \bibinfo {author} {\bibfnamefont {Y.}~\bibnamefont {Mokrousov}},
  \bibinfo {author} {\bibfnamefont {R.}~\bibnamefont {Wieser}}, \bibinfo
  {author} {\bibfnamefont {J.~E.}\ \bibnamefont {Bickel}}, \bibinfo {author}
  {\bibfnamefont {E.}~\bibnamefont {Vedmedenko}}, \bibinfo {author}
  {\bibfnamefont {S.}~\bibnamefont {Bl{\"u}gel}}, \bibinfo {author}
  {\bibfnamefont {S.}~\bibnamefont {Heinze}}, \bibinfo {author} {\bibfnamefont
  {K.}~\bibnamefont {von Bergmann}}, \bibinfo {author} {\bibfnamefont
  {A.}~\bibnamefont {Kubetzka}},\ and\ \bibinfo {author} {\bibfnamefont
  {R.}~\bibnamefont {Wiesendanger}},\ }\href@noop {} {\bibfield  {journal}
  {\bibinfo  {journal} {Phys. Rev. Lett.}\ }\textbf {\bibinfo {volume} {108}},\
  \bibinfo {pages} {197204} (\bibinfo {year} {2012})}\BibitemShut {NoStop}%
\end{thebibliography}%

\end{document}